\documentstyle[11pt,iau185,twoside,epsf]{article}

\begin{document}
\title{Confirmation of Solar-Like Oscillations in $\eta\,$Bootis}
\author{H. Kjeldsen}
\affil{Teoretisk Astrofysik Center
Aarhus Universitet, Denmark}
\author{T. R. Bedding, I. K. Baldry}
\affil{School of Physics, A28, University of Sydney, Australia}
\author{S. Frandsen, H. Bruntt, F. Grundahl, K. Lang}
\affil{Institut for Fysik og Astronomi, Aarhus Universitet, Denmark}
\author{R. P. Butler}
\affil{Carnegie Institution of Washington, USA}
\author{D. A. Fischer, G. W. Marcy}
\affil{Department of Astronomy, University of California, USA}
\author{A. Misch, S. S. Vogt}
\affil{UCO/Lick Observatory, USA}

 
\section{Excess power in $\eta$ Bootis}

Kjeldsen et al.\ (1995) detected excess power in the G0 subgiant $\eta$~Boo
from measurements of Balmer-line equivalent widths. The excess was at the
expected level, and these authors were able to extract frequency
separations and individual frequencies which agreed well with theoretical
models (Christensen-Dalsgaard et al., 1995; Guenther \& Demarque, 1996). A
more detailed discussion of theoretical models for $\eta$ Bootis was given
by Di Mauro \& Christensen-Dalsgaard (2001).

Kjeldsen et al.\ (1995) estimated the average amplitude of the strongest
modes to be 7 times solar, corresponding to 1.6 m/s in velocity. 13
individual oscillation modes were identified consistent with a large
frequency separation of 40.3 $\mu$Hz.  We note, however, that a search for
velocity oscillations in this star by Brown et al. (1997) failed to detect
a signal, setting limits at a level below that expected on the basis of the
Kjeldsen et al. result.
 
In this paper we report further observations made in 1998.  We observed this
star in Balmer-line equivalent width with the 2.5-m Nordic Optical
Telescope and in velocity with the 24-inch Lick CAT. 

\paragraph{Equivalent-width observations}

We observed $\eta$~Boo over six nights during May 1998 using ALFOSC
(Andalucia Faint Object Spectrograph and Camera) on the 2.5\,m Nordic
Optical Telescope on La Palma.  This is the same telescope used by Kjeldsen
et al.\ (1995) but with a different spectrograph.  We used seven
echelle orders covering the range 420--680\,nm with a spectral resolution
of 1700.  We obtained a total of 1843 spectra (sampling period 84\,s) in
44.2 hours over six consecutive nights (1998 May 1--6).

\paragraph{Velocity observations}

We used the Hamilton Echelle Spectrometer and the 0.6-m Coud{\'e}
Auxiliary Telescope (CAT) at Lick Observatory.  To produce high-precision
velocity measurements, the star was observed through an iodine absorption
cell mounted directly in the telescope beam.  We were allocated 56 nights
in 1998 April and May, but the weather was unseasonably poor, permitting
observations on only 26 nights (and many of these were partly lost).  On
the 11 best nights we obtained 95--120 spectra per night (sampling period
245\,s), and the total number of spectra obtained was 1989 (about one
third of the number possible with no weather losses).

In both sets of data we see an excess in the power spectrum, with
oscillation frequencies that confirm the earlier observations by Kjeldsen
et al.\ (1995).



\section{Conclusion}

\begin{itemize}
\item Observations in 1998 in a search for solar-like oscillations in
$\eta\,$Bootis confirm the earlier observations by Kjeldsen et al.\ (1995).
 
\item The large and small separations for $\eta$ Bootis measured to be
$(40.06 \pm 0.02)\,\mu$Hz and $(3.85 \pm 0.28)\,\mu$Hz, in excellent
agreement with theory (Di Mauro \& Christensen-Dalsgaard, 2001).
  
\item The amplitudes of p-modes are in good agreement with those predicted
by Kjeldsen \& Bedding (2001) using a $1/g$ scaling relation,
$g$ being the surface gravity. The measured
amplitudes are between 3 and 7 times solar (predicted value is 4.6 times
solar). The higher amplitudes observed by Kjeldsen et al.\ (1995) could
have been unusually high, reflecting the stochastic nature of the
excitation.  This may also explain why Brown et al.\ (1997) failed to
detect any signal.
 
\end{itemize}

\end{document}